\documentclass[preprint,showpacs,preprintnumbers,amsmath,amssymb]{revtex4}
\usepackage{bm}
\usepackage{graphicx}
\usepackage{floatflt,wrapfig}

\begin{document}
\title{Coupled spin-charge drift-diffusion approach for a two-dimensional electron gas with
Rashba spin-orbit coupling}

\author{V.V. Bryksin}
\affiliation{Physical Technical Institute, Politekhnicheskaya 26,
194021 St. Petersburg, Russia}

\author{P. Kleinert}
%\email[kl@pdi-berlin.de]{Your e-mail address}
\affiliation{Paul-Drude-Intitut f\"ur Festk\"orperelektronik,
Hausvogteiplatz 5-7, 10117 Berlin, Germany}

\date{\today}% It is always \today, today,
             %  but any date may be explicitly specified

\begin{abstract}
Based on kinetic equations for the density matrix, drift-diffusion
equations are derived for a two-dimensional electron gas with
Rashba spin-orbit coupling. Universal results are obtained for the
weak coupling case. Most interesting is the observation that with
increasing spin-orbit coupling strengths there is a sharp
transition between spin diffusion and ballistic spin transport.
For strong spin-orbit coupling, when the elastic scattering time
is much larger than the spin relaxation time, undamped
spin-coherent waves are identified. The existence of these
long-lived spin-coherent states is confirmed by exact analytical
results obtained from microscopic kinetic equations valid in the
ballistic regime.
\end{abstract}


\maketitle

\section{Introduction}
One of the major challenges of spintronics is the generation and
manipulation of a spin polarization by exclusively electronic
means in nonmagnetic semiconductors at room temperature. A lively
discussed phenomenon in this field is the spin-Hall effect
\cite{SCIE_1348,PRL_126603}, the theoretical description of which
is still very controversial. In the experimental analysis of this
effect \cite{SCIE_1910,PRL_047204,PRL_096605,PRL_126603}, an
electric-field induced spin accumulation is observed near the
edges of the sample. A widespread idea to understand this
phenomenon relies on the notion of a spin current oriented
transverse to the applied electric
field.~\cite{SCIE_1348,PRL_126603} This current is assumed to
induce a spin accumulation in the spin-Hall experiment. However,
this seemingly clear physical picture became the subject of
serious controversies in the
literature.~\cite{PRB_241315,PRB_245327,PRB_245305,PRL_066602,PRL_056602,PHYSE_31,PRL_076604,PRB_113305,PRB_165313}
The main problem results from the fact that the spin is not
conserved in spin-orbit coupled systems so that the definition of
a related spin current becomes ambiguous. Alternative approaches
seem to be more suitable, which avoid the identification of an
appropriate spin current. Most profound approaches of this kind
start from  a microscopic theory for the spin-density matrix
\cite{PRB_165313,PRB_045317,PRB_052407,PRB_075312,JPCM_5071,PRB_155308,PRL_076602,PRB_075366,PRB_035322}
or Keldysh Green functions \cite{PRB_245210,PRB_075306}, followed
by an analysis of its long-wavelength and low-frequency limit.
These treatments provide a complete physical description of
spin-related phenomena including accumulation, diffusion, and
relaxation of spin as well as magnetoelectric and spin-galvanic
effects.~\cite{PRB_155308} Based on this approach, it was recently
concluded that as soon as spins are polarized, the spin-orbit
interaction leads to spin precession and related spatial
oscillations of the spin accumulation.~\cite{PRB_033316} A
specific difficulty of the phenomenological drift-diffusion
approach is the formulation of appropriate boundary conditions
\cite{PRB_113309,PRB_115331} to solve the differential equations
for the position and time dependent spin and charge densities.

Another challenge in the field of spintronics refers to mechanisms
that allow long spin relaxation times. In many situations of
practical relevance, the main spin relaxation mechanism is due to
spin-orbit interaction (Dyakonov-Perel spin relaxation), which
leads to an effective internal magnetic field that causes spin
precession in the plane perpendicular to the magnetic field.
Momentum scattering on impurities randomly reorient the respective
precession axis so that an averaged spin dephasing results. From
an application point of view, the suppression of this spin
dephasing in spintronic devices is a demand of high priority. An
enhancement of the spin relaxation time has been predicted to
occur in spatial regions near the edges of a spin-polarized
stripe.~\cite{PRB_155317} A suppression of spin relaxation is also
observed in samples with an initial periodic spin
profile.~\cite{PRB_155317} The related spin-coherent standing wave
has a several times longer spin relaxation time than a homogeneous
electron spin polarization. The phase as well as the long spin
relaxation time of these novel spin waves could be important to
reading, writing, and manipulating information in spintronic
devices.

In this paper, we focus on coupled spin-charge drift-diffusion
equations, which allow a treatment of spin-related phenomena that
are interesting both from a theoretical and experimental point of
view. Starting from microscopic transport equations for the
spin-density matrix of a two-dimensional electron gas (2DEG) with
spin-orbit coupling of the Rashba type, we derive in an exact
manner drift-diffusion equations for coupled spin-charge
excitations. These equations, which are treated both for weak and
strong spin-orbit coupling, allow a due consideration of both the
field-induced spin accumulation and spin-coherent excitations. In
the weak-coupling limit, the basic results are universal, i.e.,
they qualitatively agree with findings recently obtained for
completely other systems, e.g., the spin transport of small
polarons.~\cite{PRB_235302} Whereas the weak-coupling case has
received already considerable interest in the literature
\cite{PRL_226602,PRB_155308,PRL_076602}, there are only few
considerations of effects due to strong spin-orbit coupling. As
shown below completely new and interesting phenomena are predicted
to appear. We mention the transition from the diffusive to the
ballistic spin transport regime and the occurrence of
spin-coherent standing waves. In this paper, emphasis is put on
the interesting regime of strong spin-orbit interaction.

\section{Basic theory}
Coupled spin and charge excitations are treated by an
effective-mass Hamiltonian that describes electrons and their spin
of a 2DEG with short-range spin-independent elastic scattering on
impurities. In addition, the carriers are coupled to each other
via spin-orbit interaction of Rashba type. To keep the approach
transparent, we postpone the consideration of effects due to
external electric and magnetic fields to a forthcoming
publication. The second quantized form of the Hamiltonian reads
\begin{equation}
H_{0}=\sum_{\bm{k},\lambda }a_{\bm{k}\lambda }^{\dag}\left[ \varepsilon_{%
\bm{k}}-\varepsilon _{F}\right] a_{%
\bm{k}\lambda }-\sum_{\bm{k},\lambda ,\lambda ^{\prime }}\left(
\hbar{\bm{\omega}}_{
\bm{k}} \cdot {\bm{\sigma }}_{\lambda \lambda ^{\prime }}\right) a_{\bm{k}%
\lambda }^{\dag}a_{\bm{k}\lambda ^{\prime }}
+u\sum\limits_{\bm{k},
\bm{k}^{\prime}}\sum\limits_{\lambda}a_{\bm{k}\lambda}^{\dag}a_{\bm{k}^{\prime}\lambda},
\label{Hamilton}
\end{equation}
where $a_{\bm{k}\lambda}^{\dag}$ ($a_{\bm{k}\lambda}$) denote the
creation (annihilation) operators with in-plane quasi-momentum
$\bm{k}=(k_x,k_y,0)$ and spin $\lambda$. In Eq.~(\ref{Hamilton}),
$\varepsilon_F$ denotes the Fermi energy, $\bm{\sigma}$ the vector
of Pauli matrices, and $u$ the strength of the 'white-noise'
elastic impurity scattering, which gives rise to the momentum
relaxation time $\tau$ calculated from
\begin{equation}
\frac{1}{\tau}=\frac{2\pi u^2}{\hbar}\sum\limits_{\bm{k}^{\prime}}
\delta(\varepsilon_{\bm{k}^{\prime}}-\varepsilon_{\bm{k}}).%
\label{deftau}
\end{equation}
Other quantities in Eq.~(\ref{Hamilton}) are defined by
\begin{equation}
\varepsilon_{\bm{k}}=\frac{\hbar^2\bm{k}^2}{2m},\quad
{\bm{\omega}}_{ \bm{k}}=\frac{\hbar}{m}(\bm{K}\times\bm{k}),\quad
\bm{K}=\frac{m\alpha}{\hbar^2}{\bm{e}}_z , \label{eq3}
\end{equation}
with $\alpha$ denoting the spin-orbit coupling constant and $m$
the effective mass.

The complete information about all physical quantities of interest
are included in the charge and spin components of the spin-density
matrix
\begin{equation}
f_{\lambda^{\prime}}^{\lambda}(\bm{k},\bm{k}^{\prime}\mid
t)=\langle
a^{\dag}_{\bm{k}\lambda}a_{\bm{k}^{\prime}\lambda^{\prime}}\rangle_t,
\label{eq4}
\end{equation}
which is more conveniently expressed in the following manner
\begin{equation}
f(\bm{k},\bm{\kappa}\mid
t)=\sum\limits_{\lambda}f_{\lambda}^{\lambda}(\bm{k},\bm{\kappa}\mid
t),\quad {\bm{f}}(\bm{k},\bm{\kappa}\mid
t)=\sum\limits_{\lambda,\lambda^{\prime}}f_{\lambda^{\prime}}^{\lambda}(\bm{k},\bm{\kappa}\mid
t){\bm{\sigma}}_{\lambda,\lambda^{\prime}}.%
\label{eq5}
\end{equation}
The $\bm{\kappa}$ dependence, introduced by the replacements
$\bm{k}\rightarrow \bm{k}+\bm{\kappa}/2$,
$\bm{k}^{\prime}\rightarrow \bm{k}-\bm{\kappa}/2$, refers to a
possible spatial charge and/or spin inhomogeneity and leads to a
coupling between charge- and spin-degrees of freedom. Treating
elastic impurity scattering in the Born approximation and keeping
only the lowest-order contributions of the spin-orbit interaction
in the collision integral, we obtain the following
Laplace-transformed kinetic equations \cite{PRB_165313}
\begin{equation}
sf-\frac{i\hbar}{m}(\bm{\kappa}\cdot\bm{k})f
-\frac{i\hbar}{m}\bm{K}\cdot({\bm{f}}\times\bm{\kappa})
=\frac{1}{\tau}(\overline{f}-f)+f_0, \label{kin1}
\end{equation}
\begin{eqnarray}
&&s{\bm{f}}+2({\bm{\omega}}_{\bm{k}}\times{\bm{f}})
-\frac{i\hbar}{m}(\bm{\kappa}\cdot\bm{k}){\bm{f}}
+\frac{i\hbar}{m}(\bm{K}\times\bm{\kappa})f\nonumber\\
&&=\frac{1}{\tau}(\overline{{\bm{f}}}-{\bm{f}})+\frac{1}{\tau}
\frac{\partial}{\partial\varepsilon_{\bm{k}}}
\overline{f\hbar{\bm{\omega}}_{\bm{k}}}-\frac{\hbar{\bm{\omega}}_{\bm{k}}}{\tau}
\frac{\partial}{\partial\varepsilon_{\bm{k}}}
\overline{f}+\bm{f}_0, \label{kin2}
\end{eqnarray}
where $f_0$ and $\bm{f}_0$ denote the initial charge and spin
distribution, respectively. The cross-line over $\bm{k}$-dependent
quantities denotes an integration over the angle $\varphi$ of the
in-plane vector $\bm{k}$ and $s$ is the variable of the Laplace
transformation that replaces the time $t$. The kinetic
Eqs.~(\ref{kin1}) and (\ref{kin2}) have been obtained under the
condition $\hbar^2K^2/(m\varepsilon_F)\ll 1$, when the spin
contribution on the left-hand side of Eq.~(\ref{kin1}) can be
considered as a small correction.

To proceed we derive coupled spin-charge drift-diffusion
equations, which are completely in accordance with the kinetic
Eqs.~(\ref{kin1}) and (\ref{kin2}) in the long-wavelength limit.
The Eqs.~(\ref{kin1}) and (\ref{kin2}) are written in an more
convenient and explicit form as a set of four coupled linear
equations
\begin{eqnarray}
\sigma f+i\Omega(q_xf_y-q_yf_x)&=&R\nonumber\\
\sigma f_x+2\Omega\cos(\varphi) f_z-i\Omega q_y
f&=&R_x+\Omega\sin(\varphi) \frac{\hbar}{\tau}
\frac{\partial \overline{f}}{\partial\varepsilon_{\bm{k}}}\nonumber\\
\sigma f_y+2\Omega\sin(\varphi) f_z+i\Omega q_x
f&=&R_y-\Omega\cos(\varphi) \frac{\hbar}{\tau}
\frac{\partial \overline{f}}{\partial\varepsilon_{\bm{k}}}\nonumber\\
\sigma f_z-2\Omega\left[\cos(\varphi) f_x+\sin(\varphi)
f_y\right]&=&R_z . \label{kin}
\end{eqnarray}
Here, we introduced the abbreviations
$\Omega=\omega_{\bm{k}}\tau$, $q_{x,y}=\kappa_{x,y}/k$ and
\begin{equation}
\sigma=\sigma_0-i\Omega\frac{k}{K}\left[q_x\cos(\varphi)+q_y\sin(\varphi)\right],\quad
\sigma_0=s\tau+1 .
\end{equation}
The quantities on the right-hand side of Eq.~(\ref{kin}) are given
by
\begin{equation}
R=\overline{f}+\tau f_0,\quad R_z=\overline{f}_z+\tau f_{0z},
\end{equation}
\begin{equation}
R_x=\overline{f}_x+\tau
f_{0x}-\frac{\hbar}{\tau}\frac{\partial}{\partial\varepsilon_{\bm{k}}}
\Omega \overline{f\sin(\varphi)},\quad R_y=\overline{f}_y+\tau
f_{0y}+\frac{\hbar}{\tau}\frac{\partial}{\partial\varepsilon_{\bm{k}}}
\Omega \overline{f\cos(\varphi)}.
\end{equation}
After solving the set of linear Eqs.~(\ref{kin}) for the
components of the density matrix, the solution is expanded up to
second order in $q_x$ and $q_y$. The final analytical integration
over the angle $\varphi$ results in closed equations for the
physical quantities of interest namely $\overline{f}$ and
$\overline{\bm{f}}$. The charge density $\overline{f}$, is
expressed by $\overline{f}(\varepsilon_{\bm{k}},q\mid
s)=n(\varepsilon_{\bm{k}})F(q\mid s)$, where
$n(\varepsilon_{\bm{k}})$ denotes the Fermi distribution function
for carriers in thermal equilibrium. This ansatz is fundamental
and completely in line with our basic assumption concerning the
relevant time scale of drift-diffusion effects, which should last
long enough to guarantee that carriers quickly reestablish the
thermal distribution. For these relaxation processes, the energy
relaxation time $\tau_{\varepsilon}$ is responsible. Usually, the
inelastic scattering time is much larger than $\tau$ so that the
time scale $t>\tau_{\varepsilon}$ that can be accounted for by the
widespread drift-diffusion approach is restricted to $t\gg\tau$
(or $s\tau\ll 1$). The calculation outlined above is
straightforward and leads to the result
\begin{equation}
\frac{\partial}{\partial\varepsilon_{\bm{k}}}\Omega\,\overline{f\sin(\varphi)}=
\frac{i\kappa_y}{\sigma_0}\frac{\tau^2K}{m}(\varepsilon_{\bm{k}}\overline{f})^{\prime},\qquad
\frac{\partial}{\partial\varepsilon_{\bm{k}}}\Omega\,\overline{f\cos(\varphi)}=
\frac{i\kappa_x}{\sigma_0}\frac{\tau^2K}{m}(\varepsilon_{\bm{k}}\overline{f})^{\prime},
\end{equation}
where the prime denotes the derivative with respect to
$\varepsilon_{\bm{k}}$. This result is used to derive equations
for all components of the density matrix, which decouple into two
sets of equations for the components $\overline{f}$ and
$i({\bm{\kappa}}\times{\overline{\bm{f}}})_z=\overline{f_r}$ on
the one hand and $\overline{f}_z$ and
${i\bm{\kappa}}\cdot{\overline{\bm{f}}}=\overline{f_d}$ on the
other hand. We obtain
\begin{equation}
\left[s+D_0(s)\kappa^2\right]\overline{f}+V(s)\overline{f}_r=f_0,
\label{k1}
\end{equation}
\begin{equation}
\left[ s+\frac{1}{\tau_{s\perp}(s)}+D_r(s)\kappa^2
\right]\overline{f}_r-V_r(s)\kappa^2 \overline{f}=f_{r0},
\label{k2}
\end{equation}
\begin{equation}
\left[ s+\frac{1}{\tau_{sz}(s)}+D_z(s)\kappa^2
\right]\overline{f}_z-V_z(s)\overline{f}_d=f_{z0} , \label{k3}
\end{equation}
\begin{equation}
\left[ s+\frac{1}{\tau_{s\perp}(s)}+D_d(s)\kappa^2
\right]\overline{f}_d-V_d(s)\kappa^2\overline{f}_z=f_{d0} ,
\label{k4}
\end{equation}
with the following diffusion coefficients
\begin{equation}
D_0(s)=\frac{D}{\sigma_0^2}, \qquad D=v^2\tau/2,
\end{equation}
\begin{equation}
D_r(s)=\frac{\sigma_0^6+6\sigma_0^4\Omega^2+36\sigma_0^2\Omega^4+48\Omega^6}
{\sigma_0^2(\sigma_0^2+4\Omega^2)^2(\sigma_0^2+2\Omega^2)}D,
\end{equation}
\begin{equation}
D_z(s)=\frac{\sigma_0^2-12\Omega^2}{(\sigma_0^2+4\Omega^2)^2}D,
\qquad
D_d(s)=\frac{\sigma_0^6-6\sigma_0^4\Omega^2+12\sigma_0^2\Omega^4+16\Omega^6}
{\sigma_0^2(\sigma_0^2+4\Omega^2)^2(\sigma_0^2+2\Omega^2)}D,\label{Dz}
\end{equation}
spin relaxation times
\begin{equation}
\tau_{s\perp}(s)=\tau_s\frac{\sigma_0^2+2\Omega^2}{\sigma_0},\qquad
\tau_{sz}(s)=\tau_s\sigma_0/2, \qquad
1/\tau_s=4DK^2=2\Omega^2/\tau
\end{equation}
and coupling terms
\begin{equation}
V(s)=\frac{v}{\sigma_0}\frac{K}{k}\frac{\sigma_0^2+2\Omega^2}{\sigma_0^2+4\Omega^2},\qquad
V_r(s)=\frac{v}{\sigma_0}\frac{K}{k}\left[
1+\frac{\sigma_0^2+4\Omega^2}{\sigma_0^2+2\Omega^2}
\frac{\varepsilon_{\bm{k}}n^{\prime}}{n} \right],
\end{equation}
\begin{equation}
V_z(s)=v\frac{2\Omega}{\sigma_0^2+4\Omega^2},\qquad
V_d(s)=v\frac{2\Omega\sigma_0^2}{(\sigma_0^2+4\Omega^2)(\sigma_0^2+2\Omega^2)}.
\end{equation}
The velocity $v$ is given by $\hbar k/m$. All spin relaxation
times are proportional to $1/\tau$ ($1/\tau_s=2\omega_k^2\tau$),
which is in accordance with the Dyakonov-Perel spin-relaxation
mechanism. The basic solution in Eqs.~(\ref{k1}) to (\ref{k4})
allows in principle the consideration of non-Markovian
spin-related phenomena because all transport coefficients depend
on the variable $s$ of the Laplace transformation. Moreover, in
the strong-coupling limit ($\Omega\gg 1$), when the inequality
$\tau\gg\tau_s$ is satisfied, the spin diffusion exhibits
principal a non-Markovian character. Unfortunately, it turns out
that the full consideration of the complicated time dependence in
the drift-diffusion Eqs.~(\ref{k1}) to (\ref{k4}) is not an
passable route. A discussion of this mathematical subtlety is
postponed to a forthcoming paper. In the weak-coupling limit
$\Omega\ll 1$, under the condition $\tau \ll\tau_s$, the spin
evolution proceeds completely in a Markovian manner.

A striking feature of our basic result is the decomposition of the
transport Eqs.~(\ref{k1}) to (\ref{k4}) into two sets of coupled
equations for (i) the charge density $\overline{f}$ and the
transverse spin component $\overline{f}_r$ and (ii) for
$\overline{f}_z$ and the longitudinal spin component
$\overline{f}_d$. This feature has already been identified for
spin-orbit coupled extended electronic states \cite{PRL_076602}
and for hopping transitions of small polarons.~\cite{PRB_235302}
As a consequence of this decomposition, the internal effective
magnetic field ${\bm{H}}$, which is associated with the spin-orbit
interaction, is only due to the spin components $\overline{f}_z$
and $\overline{f}_d$ but not to $\overline{f}_r$. From the Maxwell
equation, we obtain ${\bm{H}}({\bm{r}})=-\nabla \varphi
({\bm{r}})$ with
\begin{equation}
\varphi({\bm{r}})=\frac{\mu_B}{2\pi}\sum\limits_{\bm{k}}\int
d{\bm{\kappa}} \exp\left[ -i\kappa_xx-i\kappa_yy-\kappa\mid
z\mid\right]\left[\frac{\overline{f}_d({\bm{k}},{\bm{\kappa}})}{\kappa}-\frac{z}{\mid
z\mid}\overline{f}_z({\bm{k}},\kappa) \right],
\end{equation}
where $\mu_B$ denotes the Bohr magnaton. The measurable magnetic
properties are therefore completely determined by Eqs.~(\ref{k3})
and (\ref{k4}). With other words, the carrier diffusion leaves the
spin-induced magnetization unchanged. The situation changes
drastically, when an electric field is applied. In this case, all
components of the density matrix couple to each other
\cite{PRB_235302} so that an inhomogeneity in the distribution of
charge carriers may also induce a magnetization. Recently this
magneto-electric effect has been thoroughly investigated in the
literature for semiconductors with spin-orbit interaction.

\section{Weak spin-orbit coupling}
The character of spin transport and spin diffusion strongly
depends on the strength of the spin-orbit coupling, which is
expressed by the dimensionless parameter $\Omega=Kl$. In this
Section, we focus on the weak-coupling case $\Omega\ll 1$, when
the mean free path $l$ is much shorter than the characteristic
length $K^{-1}$ of the spin-orbit coupling. Accordingly, the
drift-diffusion Eqs.~(\ref{k1}) to (\ref{k4}) are expressed by
\begin{equation}
(s+D\kappa^2)\overline{f}+\frac{\hbar K}{m}f_r=f_0,\label{kk1}
\end{equation}
\begin{equation}
\left(s+\frac{1}{\tau_s}+D\kappa^2 \right)\overline{f}_r
-\frac{\hbar K}{m}\frac{(\varepsilon_{\bm{k}}
n)^{\prime}}{n}\kappa^2\overline{f}=f_{r0},\label{kk2}
\end{equation}
\begin{equation}
\left(s+\frac{2}{\tau_s}+D\kappa^2 \right)\overline{f}_z
-4DK\overline{f}_d=f_{z0},\label{kk3}
\end{equation}
\begin{equation}
\left(s+\frac{1}{\tau_s}+D\kappa^2 \right)\overline{f}_d
-4DK\kappa^2\overline{f}_z=f_{d0},\label{kk4}
\end{equation}
with the spin relaxation time
\begin{equation}
\tau_s=\frac{\tau}{2\Omega^2}=\frac{1}{4DK^2}.\label{taus}
\end{equation}
These equations formally agree with results obtained for the
hopping transport of small polarons.~\cite{PRB_235302} The
similarity becomes complete, when the Hall mobility $u_H$ in the
latter approach is replaced by the quantity
$(e\tau_s/m)(\varepsilon_{\bm{k}} n)^{\prime}/n$. Therefore, we
conclude that independent of the transport mechanism, the kinetic
Eqs.~(\ref{kk1}) to (\ref{kk4}) are universal for weakly
spin-orbit coupled carriers. This conclusion is confirmed by
similar studies in the literature.\cite{PRL_226602,PRB_155308}

The weak-coupling limit is most important for the hopping regime,
where the hopping length must be identified with the mean free
path. For the transport of small polarons, this length is the
lattice constant $a$ so that the weak-coupling condition $Ka\ll 1$
is almost always satisfied.

Another conclusion follows directly from Eq.~(\ref{taus}). For
weak spin-orbit coupling, the spin relaxation time $\tau_s$ is
much larger than $\tau$. Consequently, the charge degree of
freedom approaches more quickly the state of equilibrium than the
spin degree of freedom so that the frequency response of
spin-related quantities to external perturbations does not exhibit
a resonance character. According to Eq.~(\ref{kk2}), an evolution
of the transverse spin component $\overline{f}_r$ is expected that
follows adiabatically the time variation of $\overline{f}$ when
$\tau_s<\tau_{\varepsilon}$.

The solution of the set of Eqs.~(\ref{kk1}) to (\ref{kk4}) has
already been thoroughly studied previously.~\cite{PRB_155308} From
the coupled Eqs.~(\ref{kk1}) and (\ref{kk2}), an interesting
spin-related effect has been predicted in the literature
\cite{PRL_076602} namely the spontaneous splitting of a density
packet into two counter propagating packets with opposite spins.
Other results, which are obtained from the universal
Eqs.~(\ref{kk1}) to (\ref{kk4}) were derived and discussed in
Ref.~[\onlinecite{PRB_235302}].

Finally, let us add a remark concerning the component of the
spin-density vector ${\overline{\bm{f}}}$, which is proportional
to ${\bm{\kappa}}$. To lowest and first-order in ${\bm{\kappa}}$,
we obtain for the charge density
$\overline{f}=n(\varepsilon_{\bm{k}})/s$. Using this result and
accounting for Eqs.~(12), (\ref{k2}), and (\ref{k4}), we arrive at
a solution
\begin{equation}
\overline{\bm{f}}_{\bm{\kappa}}=\frac{-i{\bm{\omega_{\bm{\kappa}}}}\tau}{\sigma_0s}
\left[
(\varepsilon_{\bm{k}}n)^{\prime}-\frac{2\Omega^2\sigma_0n}{\sigma_0^2s\tau
+2\Omega^2(2s\tau+1)} \right]
\end{equation}
that was derived already in a previous paper.~\cite{PRB_165313}
This particular contribution gives rise to a dissipationless spin
current in the ground state \cite{PRB_165313} that reflects the
time evolution of the spin accumulation. Due to the pole structure
of this expression, resonances are predicted to occur at
characteristic frequencies.~\cite{PRB_165313,PRL_226602}

\section{Strong spin-orbit coupling}
For strong spin-orbit coupling $\Omega\gg 1$, the character of the
set of Eqs.~(\ref{k1}) and (\ref{k2}) for $\overline{f}$ and
$\overline{f}_r$ does not change remarkably. We obtain $D_r=3D/2$
and $\tau_{s\perp}=\tau$, and therefore
$\tau_{s\perp}\ll\tau_{\varepsilon}$ so that in the regime
$t\gg\tau$, which is accessible by the drift-diffusion approach,
the transverse spin component $\overline{f}_r$ adiabatically
follows the evolution of the charge density $\overline{f}$. To
confirm this assertion, we analyze Eq.~(\ref{k2}) in the
strong-coupling limit. Neglecting the small contributions $s$ and
$3D\kappa^2/2$ on the left-hand side of Eq.~(\ref{k2}), we
immediately obtain
\begin{equation}
\overline{f}_r(\varepsilon_{\bm{k}},{\bm{r}},t))=\frac{\hbar
K\tau}{mD}\left(1+\frac{\varepsilon_{\bm{k}} n^{\prime}}{n}
\right)\frac{\partial}{\partial
t}\overline{f}(\varepsilon_{\bm{k}},{\bm{r}},t),
\end{equation}
where $\overline{f}(\varepsilon_{\bm{k}},{\bm{r}},t)$ denotes the
exact solution of Eq.~(\ref{k1}) with $V(s)=0$.

While the set of Eqs.~(\ref{k1}) and (\ref{k2}) for the quantities
$\overline{f}$ and $\overline{f}_r$ provide expected solutions
also in the strong coupling limit, we observe a dramatic and
abrupt change in the behavior of the spin components
$\overline{f}_z$ and $\overline{f}_d$ described by Eqs.~(\ref{k3})
and (\ref{k4}). As seen from Eq.~(\ref{Dz}), the diffusion
coefficient $D_z$ vanishes at $\Omega=1/\sqrt{12}$ so that for
this coupling strength the physical picture of diffusion
completely collapses. With further increasing $\Omega$, the
diffusion coefficient $D_z$ becomes negative, which indicates an
instability in the spin subsystem implying that the applicability
of the kinetic Eqs.~(\ref{k3}) and (\ref{k4}) becomes problematic.
According to Eq.~(\ref{taus}), we have $\tau_s<\tau$ in the
strong-coupling limit ($\Omega>1$). Consequently, the spin
diffusion evolves on a time scale that is shorter than the elastic
scattering time $\tau$ so that the kinetic Eqs.~(\ref{k3}) and
(\ref{k4}) describe in principle a non-Markovian behavior.
Consequently, only in the steady state ($s\rightarrow 0$), one can
strictly speak about a Markovian spin diffusion in this regime. As
shown below, another peculiarity arises namely that the quantities
$\overline{f}_z$ and $\overline{f}_d$ rapidly change on a length
scale of the order of the mean free path $l$, a result which does
not correspond to the ${\bm{\kappa}}$ expansion of
Eqs.~(\ref{kin1}) and (\ref{kin2}). Putting all together, we
conclude that with increasing coupling strength $\Omega$, a sharp
transition appears from a diffusive behavior to a ballistic
regime. This transition could be due both to an increasing
spin-orbit coupling $K$ and an increasing relaxation time $\tau$.

Since the principal applicability of the drift-diffusion approach
for strongly spin-orbit coupled carriers becomes doubtful, we
return to the basic Eqs.~(\ref{kin1}) and (\ref{kin2}) and
consider its exact solutions, which are valid in the pure
ballistic regime ($\tau\rightarrow\infty$) that corresponds to the
extreme strong coupling limit. First, we treat the ballistic
evolution of a spin packet that is initially ($t=0$) created at
${\bm{r}}={\bm{0}}$. In the limit $\tau\rightarrow\infty$, the
kinetic Eqs.~(\ref{kin1}) and (\ref{kin2}) are solved by
\begin{equation}
f_z({\bm{\kappa}},s)=\frac{f_{z0}}{\beta({\bm{\kappa}},s)
+4\omega_{\bm{k}}^2/\beta({\bm{\kappa}},s)}, \qquad
\beta({\bm{\kappa}},s)=s-i({\bm{\kappa}}\cdot{\bm{v}}),
\label{ee1}
\end{equation}
where ${\bm{v}}=\hbar{\bm{k}}/m$ denotes the electron velocity and
where an additional contribution has been omitted that disappears
after the integration over the angle $\varphi$. The back
transformation to the ${\bm{r}}$, $t$ representation is easily
accomplished and we get
\begin{equation}
\overline{f}_z({\bm{r}},t)=f_{z0}\cos(2Kr)\delta({\bm{r}}^2-{\bm{v}}^2t^2).
\label{ee2}
\end{equation}

This solution expresses the pure ballistic character of the spin
transport.
%======================================================================
%%Fig.1
\begin{floatingfigure}{7.0cm}
\centerline{\includegraphics*[width=7.0cm]{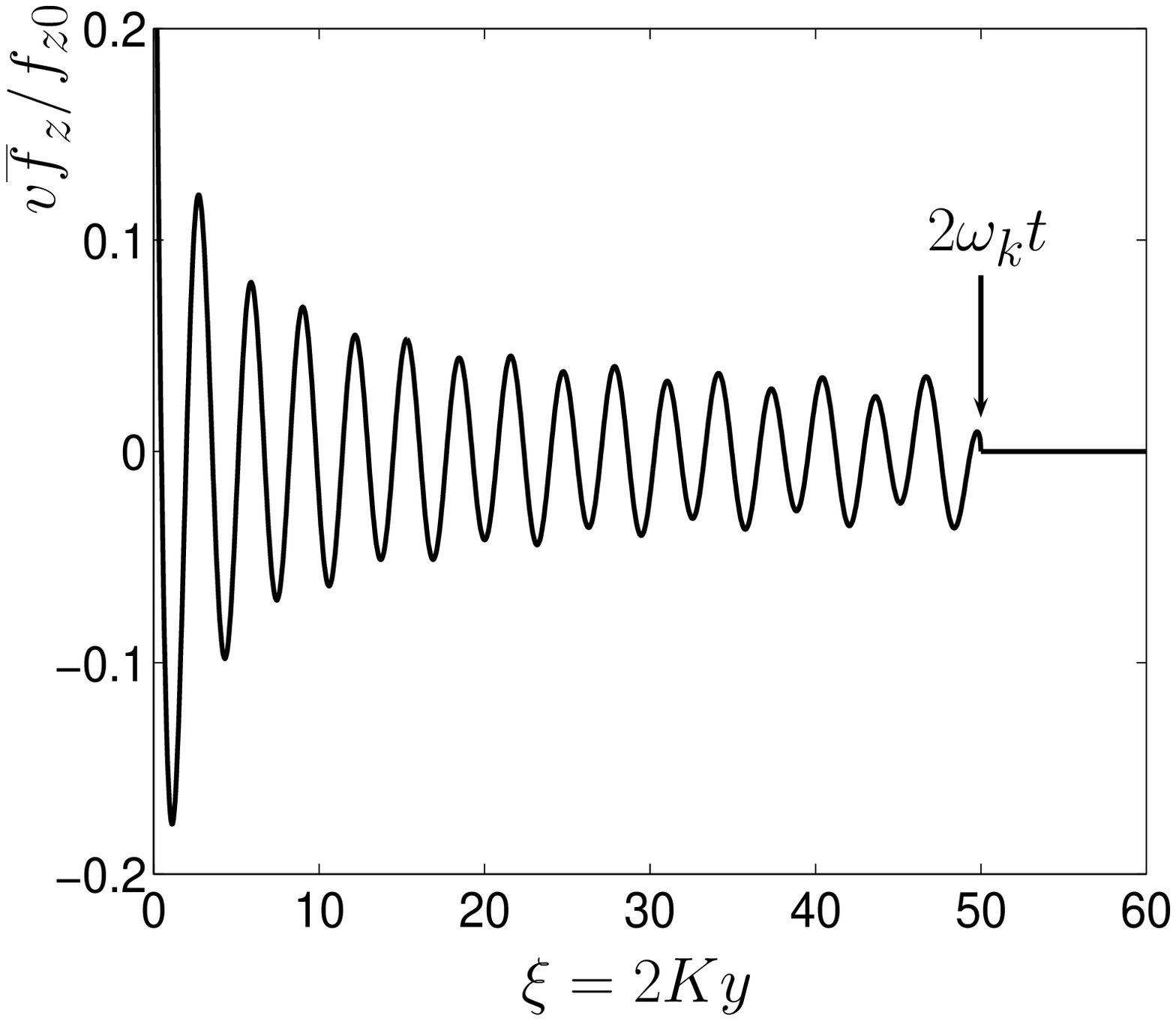}}%
\caption{\footnotesize\baselineskip=12pt Spin polarization $v
\overline{f}_z/f_{z0}$ as a function of $\xi=2Ky$ calculated for
the half space $y>0$ from Eq.~(33). \label{figure1}}
\end{floatingfigure}
%======================================================================
The front of the conic spin packet moves with the constant
velocity $v$ whereas its amplitude oscillates with a frequency
$2Kv$ due to spin rotation. This oscillation manifests itself by a
spatial periodic pattern, which in the time domain describes spin
precession with the Lamor frequency $2\omega_k$.%\\
Let us treat another example namely the half plane $y>0$ with a
given spin polarization at the boundary $y=0$. This steady-state
boundary condition is described by a term $f_{z0}/s$ on the
right-hand side of Eq.~(\ref{kin2}). Again, we start from the
solution in Eq.~(\ref{ee1}), restrict to the $y$ dependence and
perform the inverse Fourier and Laplace transformations. The
solution
\begin{equation}
\frac{\overline{f}_z}{f_{z0}}=\frac{2\omega_{\bm{k}}t-\xi}{2\pi v}
\Theta(2\omega_{\bm{k}}t-\xi)\int\limits_{0}^{1/\xi} d\eta
\frac{\cos \left[((2\omega_{\bm{k}}t-\xi)\eta +1)
\right]}{\sqrt{((2\omega_{\bm{k}}t-\xi)\eta +1)^2-1}},
\end{equation}
manifests a spin front moving with velocity $v$ in the $y$
direction ($\xi =2Ky$). In contrast to the point source in
Eq.~(\ref{ee2}), one observes non-vanishing oscillatory spin
excitations also in the interior. Spin-coherent waves are formed
as shown in Fig.~\ref{figure1}. The physical origin of this spin
pattern is due to the instability of the system expressed by a
negative diffusion coefficient ($D_z<0$). In this regime, spin
diffusion has the tendency to strengthen an initial spin
inhomogeneity, whereas the competition with spin relaxation leads
to the spatial oscillatory spin pattern.

We point out that the exact solutions of the kinetic
Eqs.~(\ref{kin1}) and (\ref{kin2}) in the ballistic regime
describe an oscillation spin pattern, which changes on a length
scale $K^{-1}$ that is much smaller than the mean free path $l$
($\Omega=Kl\gg 1$). These exact results derived for the ballistic
regime will help to recognize the solution of the drift-diffusion
Eqs.~(\ref{k3}) and (\ref{k4}) in the limit of strong spin-orbit
coupling.

We start our analysis of these equations by calculating its
possible eigenmodes. Treating the determinant of Eqs.~(\ref{k3})
and (\ref{k4}) in the limit $s\rightarrow 0$, we see that there is
no oscillatory mode in the weak-coupling regime
$\Omega<1/(3\sqrt{3})$. However, for coupling strengths in the
interval $1/(3\sqrt{3})<\Omega <1/(2\sqrt{3})$, there are two
solutions with vanishing imaginary parts that describe completely
undamped oscillations. For even higher spin-orbit interaction
$1/(2\sqrt{3})<\Omega$, only one oscillatory solution is obtained.
To illustrate this behavior let us treat a stripe of width $2L$
parallel to the $x$ axis ($-L\le y\le L$). A permanent spin
polarization should be provided at the boundaries
[$\overline{f}_z(y=\pm L, s)=f_{z0}/s$]. For this special case,
the inverse Fourier transformation of Eqs.~(\ref{k3}) and
(\ref{k4}) leads to the following dimensionless differential
equations
\begin{equation}
d_z\overline{f}_z^{\prime\prime}-\alpha_z\overline{f}_z+\gamma_dF_d=0,
\end{equation}
\begin{equation}
d_dF_d^{\prime\prime}-\alpha_dF_d-\gamma_z\overline{f}_z^{\prime\prime}=0,
\end{equation}
with $F_d=\overline{f}_d/K$. The derivative refers to $\xi =2Ky$
and the parameters are given by
\begin{equation}
\alpha_z=\sigma_0s\tau+4\Omega^2,\quad
\alpha_d=\sigma_0^2s\tau+2\Omega^2(2s\tau+1),
\end{equation}
\begin{equation}
\gamma_z=8\Omega^2\frac{\sigma_0^2}{\sigma_0^2+4\Omega^2},\quad
\gamma_d=\frac{2\sigma_0\Omega^2}{\sigma_0^2+4\Omega^2},
\end{equation}
\begin{equation}
d_z=2\sigma_0\Omega^2\frac{\sigma_0^2-12\Omega^2}{(\sigma_0^2+4\Omega^2)^2},\quad
d_d=2\Omega^2\frac{\sigma_0^6-6\sigma_0^4\Omega^2+12\sigma_0^2\Omega^4+16\Omega^6}
{\sigma_0^2(\sigma_0^2+4\Omega^2)^2}.
\end{equation}
With respect to the $\xi$ dependence, the analytic solution of
these equations is easily derived. For the boundary conditions
$f_{z0}\ne 0$ and $f_{y0}=0$, we obtain
\begin{equation}
\overline{f}_z=\frac{f_{z0}}{s}
\frac{\kappa_1(\alpha_z-d_z\kappa_2^2)\sinh\kappa_2\xi_0\cosh\kappa_1\xi
-\kappa_2(\alpha_z-d_z\kappa_1^2)\sinh\kappa_1\xi_0\cosh\kappa_2\xi}
{\kappa_1(\alpha_z-d_z\kappa_2^2)\sinh\kappa_2\xi_0\cosh\kappa_1\xi_0
-\kappa_2(\alpha_z-d_z\kappa_1^2)\sinh\kappa_1\xi_0\cosh\kappa_2\xi_0},
\label{e39}
\end{equation}
with $\xi_0=2KL$. $\kappa_{1,2}$ are solutions of the biquadratic
equation
\begin{equation}
(d_z\kappa^2-\alpha_z)(d_d\kappa^2-\alpha_d)+\gamma_z\gamma_d\kappa^2=0,
\end{equation}
with the plus sign. The remaining inverse Laplace transformation
requires a numerical integration.

First, let us treat the steady-state solution ($s=0$) of
Eq.~(\ref{e39}).
%======================================================================
%%Fig.2
\begin{floatingfigure}{7.0cm}
\centerline{\includegraphics*[width=7.0cm]{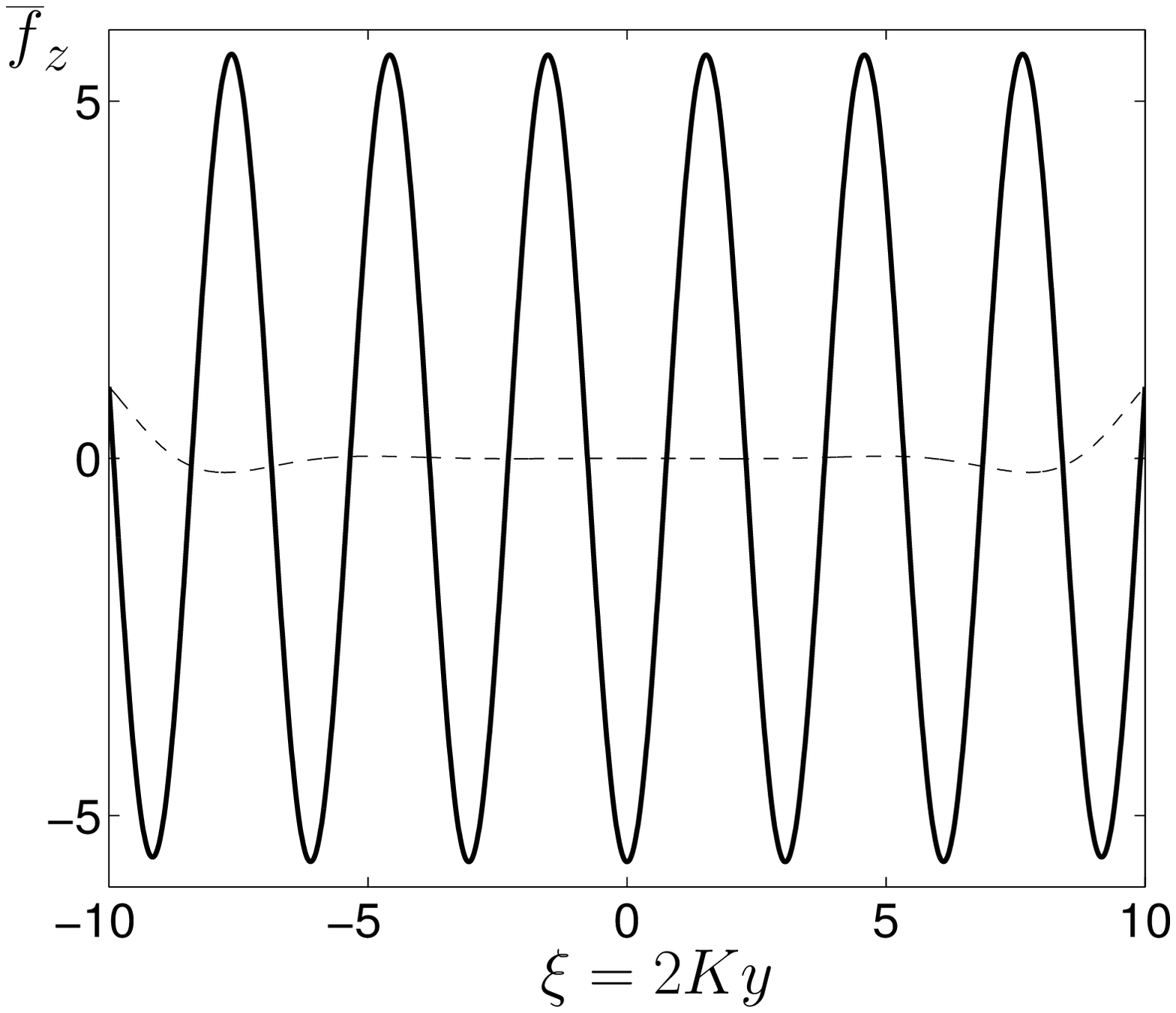}}%
\caption{\footnotesize\baselineskip=12pt Spin polarization $v
\overline{f}_z$ as a function of $\xi=2Ky$ for $\Omega=0.1$
(dashed line) and $\Omega=1$ (solid line). \label{figure2}}
\end{floatingfigure}
%======================================================================
Figure~2 shows the spin polarization $\overline{f}_z$ as a
function of $\xi$. As expected, for weak spin-orbit coupling
(dashed line), the spin polarization gradually decreases into the
interior of the stripe and vanishes completely at the center. A
completely other behavior exhibits the solid line calculated for
$\Omega=1$. A spin-coherent standing wave appears for strong
spin-orbit coupling. The period of this pattern is comparable with
the mean-free path $l$ ($\Omega=Kl=1$). Such rapid oscillations
are in principle beyond the accessibility of the drift-diffusion
approach. However, as already mentioned above, similar
oscillations appear also in the ballistic regime. Moreover,
recently it has been stated that the macroscopic transport
equations work well even when the spin-diffusion length is
comparable to the mean-free path.~\cite{PRB_212410} \\In the
strong-coupling limit $\Omega\gg 1$, the solution in
Eq.~(\ref{e39}) simplifies to
\begin{equation}
\overline{f}_z(k,y,s)=\frac{f_{z0}}{s}\frac{\cos(4\Omega\xi/\sqrt{6})}
{\cos(4\Omega\xi_0/\sqrt{6})}. \label{e40}
\end{equation}
The period of this oscillation [given by $l\sqrt{6}\pi/(4(Kl)^2)$]
is much smaller than the mean-free path $l$. In addition, the spin
polarization $\overline{f}_z$ in Eq.~(\ref{e40}) diverges, when
the half width $L$ of the stripe satisfies the resonance condition
$(2n+1)\pi/2=4\Omega\xi_0/\sqrt{6}$ with $n$ being any integer.
The appearance of these divergencies indicates that elastic
scattering alone cannot always balance the tendency of mutual spin
alignment in the strong spin-orbit coupling regime. To avoid this
defect of the approach, obviously both elastic and inelastic
scattering have to be treated on a full microscopic level. \\To
give a rough idea of the temporal evolution of spin-coherent
standing waves, we treat
%======================================================================
%%Fig.3
\begin{floatingfigure}{7.0cm}
\centerline{\includegraphics*[width=7.0cm]{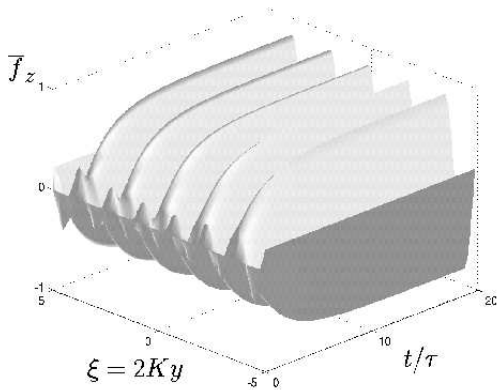}}%
\caption{\footnotesize\baselineskip=12pt Spin polarization
$\overline{f}_z$ as a function of $t/\tau$ and $\xi=2Ky$ for
$\Omega=2$ and $f_{z0}=0.2$. \label{figure3}}
\end{floatingfigure}
%======================================================================
Eq.~(\ref{e39}) in the quasi-stationary regime ($s\tau\ll 1$,
i.e., $\sigma_0=1$) and perform the inverse Laplace transformation
by a numerical integration. Figure~3 shows an example for
$\overline{f}_z$ as a function of $\xi$ and $t/\tau$. Initially,
at $t=0$, there is only the provided spin polarization $f_{z0}$ at
the boundaries. The time evolution starts with an intricate
dependence that is by no means correctly reproduced by the
drift-diffusion approach because it appears at a time scale $t
\lesssim \tau$. After this short period of rapid spatio-temporal
changes, the periodic spin pattern gradually emerges. Whereas the
gross features of this picture are certainly reliable, a detailed
analysis of the time evolution requires an approach based on a
microscopic theory.

\section{Summary}
Starting from microscopic quantum-kinetic equations for the
spin-density matrix, we derived drift-diffusion equations for a
2DEG with Rashba spin-orbit interaction. The approach is valid in
the weak as well as strong coupling regime. In the absence of
external electric and magnetic fields, the set of equations
decouples. The first set of equations relates the charge density
to the transverse in-plane spin component. Another set of
equations links the out-of plane spin component with the
longitudinal in-plane spin contribution. Whereas both sets of
equations have an universal character in the weak scattering
limit, they exhibit interesting peculiarities for strong
spin-orbit coupling. In the latter case, the spin relaxation time
$\tau_s$ becomes shorter than the elastic scattering time $\tau$.
Consequently, spin effects have a principal non-Markovian
character in this regime. Moreover, the diffusion coefficient
$D_z$ of the $\overline{f}_z$-$\overline{f}_d$ spin channel
changes its sign with increasing spin-orbit coupling $\Omega=Kl$.
This dependence gives rise to an instability in the spin subsystem
($D_z=0$) and to an undamped oscillatory spin pattern ($D_z<0$).
The wavelength of the long-lived coherent spin rotation is smaller
than the mean-free path. This fact conflicts with basic
assumptions of the drift-diffusion approach, which is only
applicable for times much longer than the elastic scattering time
and for diffusion lengths much larger than the mean-free path. The
approach signals the existence of an instability similar to the
recently studied transition from a uniform to a nonuniform ground
state in the Rashba model.~\cite{PRB_165309} The rapid spatial
variations of the out-of plane spin polarization gives rise to
microscopic circulating currents via the induced magnetic field.
Within a self-consistent schema, one expects that the retroaction
of these currents on spins may lead to a stabilization of the spin
subsystem and to a finite damping of spin oscillations.

In order to appreciate the unusual results obtained from the
drift-diffusion equations in the strong-coupling limit, an exact
treatment of the original quantum-kinetic equations is desirable.
For strong spin-orbit coupling ($\Omega\gg 1$), the spin
relaxation time is much smaller than the scattering time.
Therefore, most interesting is the consideration of the ballistic
spin regime. Exact analytical solutions of the kinetic equations
confirm the existence of a long-lived oscillating spin pattern.
This confirmation strongly suggests that spin-coherent standing
waves in a semiconducting stripe are indeed serious solutions of
the drift-diffusion equations. Moreover, a recent alternative
approach based on Monte-carlo simulations also arrived at the
conclusion that spin-coherent standing waves have an extremely
long spin-relaxation time.~\cite{PRB_155317}

We conclude that the prediction of an oscillatory spin pattern in
semiconductors with strong spin-orbit coupling of the Rashba type
seems to be interesting both from a theoretical and experimental
point of view. Further progress is expected from a due treatment
of the quantum-kinetic equations for the density matrix beyond the
pure ballistic regime ($\tau\rightarrow\infty$). In addition,
external electric and magnetic fields sensitively influence spin
coherent waves by mixing all components of the density matrix. The
experimental verification of the predicted long-lived
spin-coherent waves could be useful for future spintronic
applications.

%%%%%%%%%%%%%%%%%%%%%%%%%%%%%%%%%%%%%%%%%%%%%%%%%%%%%%%%%%%%%%%%
\begin{acknowledgments}
Partial financial support by the Deutsche Forschungsgemeinschaft
and the Russian Foundation of Basic Research under the grant
number 05-02-04004 is gratefully acknowledged.
\end{acknowledgments}
%%%%%%%%%%%%%%%%%%%%%%%%%%%%%%%%%%%%%%%%%%%%%%%%%%%%%%%%%%%%%%%%%%%
% Appendix
%\appendix
%\section{Solution of the kinetic equations}

%\bibliographystyle{prsty}
%\bibliography{abbrev,spin}

%%%%%%%%%%%%%%%%%%%%%%%%%%%%%%%%%%%%%%%%%%%%%%%%%%%%%%%%%%%%%%%%%%%

\end{document}